\documentclass[reprint,aps,showkeys]{revtex4-2}
\usepackage{graphicx}
\graphicspath{ {./images/} }
\usepackage[table]{xcolor}
\usepackage{amsmath}
\usepackage{amssymb}
\usepackage{dcolumn}
\usepackage{bbold}
\usepackage{braket}
\usepackage{lipsum} 
\usepackage{caption}
\usepackage{subcaption}
\usepackage{amsthm}
\usepackage{comment}
\usepackage{setspace}
\usepackage{hyperref}
\definecolor{darkblue}{RGB}{0,0,149}
\hypersetup{
    colorlinks=true,
    linkcolor=darkblue,
    filecolor=darkblue,
    urlcolor=darkblue,
    citecolor=darkblue,
    linktocpage=true,
}
\usepackage{url}

\def\A{A} 
\def\B{B}

\def\L{L}
\def\X{X}

\def\J{J}
\def\Ae{{\bf A}} 
\def\Be{{\bf B}}
\def\Le{{\bf L}}
\def\Xe{{\bf X}}

\def\Je{{\bf J}}
\def\Ai{{C_A}} 
\def\Bi{{C_B}}

\def\Li{{C_L}}
\def\Xi{{C_X}}

\def\Ji{{C_J}}

\begin{document}

\preprint{APS/123-QED}

\title{Quantum generalisation of Einstein's Equivalence Principle can be verified with entangled clocks as quantum reference frames}

\author{Carlo Cepollaro}
\thanks{Corresponding author}
\affiliation{
 Quantum Technology Lab, Dipartimento di Fisica, Universit\`a degli Studi di Milano, I-20133
Milano, Italy
}
\affiliation{%
	Vienna Center for Quantum Science and Technology (VCQ), Faculty of Physics, University of Vienna, Boltzmanngasse 5, A-1090 Vienna, Austria
}
\affiliation{%
	Institute of Quantum Optics and Quantum Information (IQOQI), Austrian Academy of Sciences, Boltzmanngasse 3, A-1090 Vienna, Austria
}
\email{carlo.cepollaro@oeaw.ac.at}

\author{Flaminia Giacomini}
\affiliation{%
    Perimeter Institute for Theoretical Physics, 31 Caroline St. N, Waterloo, Ontario, N2L 2Y5, Canada
}%
\affiliation{Institute for Theoretical Physics, ETH Z{\"u}rich, 8093 Z{\"u}rich, Switzerland
}
\email{fgiacomini@phys.ethz.ch}

\begin{abstract}
The Einstein Equivalence Principle (EEP) is of crucial importance to test the foundations of general relativity. When the particles involved in the test exhibit quantum properties, it is unknown whether this principle still holds. A violation of the EEP would have drastic consequences for physics. A more conservative possibility is that the EEP holds in a generalised form for delocalised quantum particles. 
Here we formulate such a generalised EEP by extending one of its paradigmatic tests with clocks to quantum clocks that are in a quantum superposition of positions and velocities. We show that the validity of such a generalised version of the EEP is equivalent to the possibility of transforming to the perspective of an arbitrary \emph{Quantum Reference Frame} (QRF), namely a reference frame associated to the quantum state of the clock. We further show that this generalised EEP can be verified
by measuring the proper time of entangled clocks in a quantum superposition of positions in the Earth gravitational field. The violation of the generalised EEP corresponds to the impossibility of defining dynamical evolution in the frame of each clock, and results in a modification to the probabilities of measurements calculated in the laboratory frame. Hence, it can be verified experimentally, for instance in an atom interferometer.

\end{abstract}

\maketitle

\section{Introduction}\label{sec:introduction}
The Einstein Equivalence Principle (EEP) is one of the cornerstones of general relativity. It is widely expected that the general-relativistic formulation of the EEP requires to be modified when either matter or gravity is quantum. However, the status of the EEP in conjunction with quantum theory is still debated: several proposals to extend it to the quantum regime have been advanced, while some authors have argued that the EEP is incompatible with situations when either matter (i.e. the particles used to test it) or gravity acquire quantum properties~\cite{aharonov1973quantum, penrose1996gravity, penrose2014gravitization, will_LR, lammerzahl1996equivalence, viola1997testing, Dimopoulos:2008hx, donoghue2015qed,donoghue2023quantum, seveso2017does, zych_QEEP, anastopoulos2018equivalence, giacomini_QRF, hardy2018, hardy2020implementation, tino2020precision, giacomini_EEP}. A physical theory in which the EEP does not hold would drastically modify our understanding of gravity, hence it is crucial to understand whether, at least in some reasonable situations in which gravity is still classical, the EEP holds in a generalised form. For instance, this could be the case for quantum particles in a quantum superposition in the gravitational field of the Earth.

A classical test of the EEP is usually performed by testing three different properties of classical test particles in the gravitational field of the Earth. In particular (see Ref.~\cite{will_LR} for an in-depth discussion), it involves a test of 1) the Weak Equivalence Principle (WEP), the local equivalence between effects due to a uniform gravitational field and those due to an accelerated frame; 2) the Local Lorentz Invariance (LLI), the independence of the outcome of the experiment from the velocity of the freely-falling frame inertial to the test particle; and 3) the Local Position Invariance (LPI), the independence of the outcome of an experiment from the position of the test particle.

When the test particles are quantum, they can be, e.g.\,in a quantum superposition of positions and velocities. A straight application of the EEP to this scenario would necessarily lead to a violation of the EEP because this would entail defining a single local reference frame associated to a system that is in superposition of different positions or velocities, and such a classical reference frame does not exist. However, this violation is not related to any gravitational effect in the theory, but to the fact that we are enforcing properties of classical particles on a quantum system: the classical EEP can only be applied in an arbitrarily small regions of phase-space, while quantum objects have a fundamental indeterminacy, due to which they cannot be arbitrarily localized. Hence, to show that a suitable extension of the EEP is still valid when tested with delocalised quantum particles, we need to adapt each aspect of the classical test of the EEP to the motion of quantum particles in a classical gravitational field. This question is experimentally relevant given that the wave-packet separation in atom interferometers is now large enough to be sensitive to the variation of the gravitational potential~\cite{Overstreet:2021hea}. In addition, quantum clocks can now discriminate differences in the gravitational redshift at the millimetre scale~\cite{bothwell2022resolving, zheng2023lab}.

In this paper, we formulate a generalised version of the EEP that holds for quantum particles in a quantum superposition state moving in the classical gravitational field of the Earth, which we call \emph{Einstein Equivalence Principle for Quantum Reference Frames}, or EEP for QRFs for short. Intuitively, this principle states that, when the quantum particles are in a quantum superposition state in a classical gravitational field, each amplitude obeys the usual version of the EEP, namely the metric is locally flat in each amplitude. This identifies a \emph{superposition of locally inertial frames} which leads to a new formulation of the EEP. Note that the formulation of this principle does not necessarily require us to consider situations in which the gravitational field is quantum: in the laboratory frame, quantum particles evolving in a classical gravitational field suffice. Our result is also compatible with an effective field theory of gravity in the quantum regime~\cite{Donoghue:1994dn, Burgess:2003jk, donoghue2023quantum}.

Key to our generalisation is the possibility to transform to the reference frame associated to such quantum particles, namely a Quantum Reference Frame (QRF). In particular, the most suitable formulation for our purposes is the one developed in Ref.~\cite{giacomini_SQRF} (see also related work~\cite{giacomini_QRF, perspective2, Hardy:2018kbp, zych2018relativity,  hoehn2018switch, hohn2019switching, giacomini2019relativistic, hoehn2019trinity, castro-ruiz, hoehn2020equivalence, streiter2020relativistic, ballesteros2020group, yang2020switching, de2020quantum, krumm2020, tuziemski2020decoherence, giacomini_EEP, giacomini_SQRF, giacomini2021quantum, mikusch2021transformation, hoehn2021quantum, hoehn2021internal, castro2021relative, de2021perspective, Overstreet:2022zgq, kabel2023quantum,delahamette2022quantum,apadula2022quantum,kabel2023quantumboundary,glowacki2023operational,hoehn2023quantum,hausmann2023measurement,de_la_Hamette_2022,delahamette2023quantum}). 

We define the EEP for QRFs by generalizing each of the three aspects of the classical EEP:

\noindent (\textbf{Q-WEP}) \emph{The local effects of (quantum) motion in a superposition of uniform gravitational fields are indistinguishable from those of an observer in flat spacetime that undergoes a quantum superposition of accelerations~\cite{giacomini_QRF}.
}

\noindent (\textbf{Q-LLI}) \emph{The outcome of any local nongravitational experiment is independent of the velocity of the
freely falling quantum reference frame in which it is performed.}

\noindent (\textbf{Q-LPI}) \emph{The outcome of any local nongravitational experiment is independent of the position of the quantum reference frame in which it is performed.}

The generalisation of the WEP that we refer to as Q-WEP is equivalent to the one in Ref.~\cite{giacomini_QRF}, where it was shown that a situation in which a quantum particle is subject to a superposition of uniform and constant accelerations is equivalent to the particle being in a quantum superposition of uniform and constant gravitational fields. In Ref.~\cite{giacomini_EEP} it was instead proposed that the EEP can be extended to quantum particles in a quantum superposition of spacetimes, i.e.\,a superposition state of the gravitational field such that classical gravity holds in each amplitude. The goal of these extended formulations of the Equivalence Principle in QRFs is to show that physical laws are not only valid for reference frames connected by standard coordinate transformations, but also for those connected by quantum superpositions of coordinate transformations, i.e.\, those connecting different QRFs. So far, however, no proposal exists about how such a principle would be tested. Here, we formulate for the first time a test of the EEP for QRFs.

In particular, we show how such EEP for QRFs can be verified in an interferometric test involving entangled quantum clocks~\cite{zych,Zych_2016,roura,roura2020,ufrecht,dipumpo,buoninfante,rosi2017,wood2021}, in the limit of weak gravitational field of the Earth and $v\ll c$, with $c$ being the speed of light. The relative phase accumulated by the clock in the two paths of the interferometer depends on the difference in proper time measured by the clock in each path. The formulation of QRFs of Ref.~\cite{giacomini_SQRF} allows us to define a proper time for such clocks in superposition, and describe the dynamical evolution of the physical systems involved in the test with respect to that proper time. A violation of the EEP for QRFs results in a modification of the proper time of the clocks associated to the particles in a quantum superposition state. This has a quantifiable effect in the probabilities measured in the interferometric test. To quantify the violation, we introduce a model for violation of the EEP for QRFs, generalising the classical one. 

The rest of the paper is organized as follows. Section~\ref{sec:GTD} is dedicated to a brief overview of atomic clocks in atom interferometers, where we review the well-known fact that atomic clocks in superposition of different paths can accumulate a relative phase that depends on the proper time difference between the paths. Hence, they can be used to measure joint effects of quantum theory and special- and general-relativistic time dilation. In Section~\ref{sec:model} we quantify the predictions of the EEP for QRFs, introducing a model for violations of the principle, and show that atomic clock interferometers can be used to test the principle. In Section~\ref{sec:QRF_perspective} we describe the predictions of the EEP for QRFs in the perspective of one of the two superposed clock, i.e. we cast the principle in terms of the proper time of a superposed clock, employing the QRFs formalism. Finally, in Section \ref{sec:test_LZ}, we show that the violation of the principle leads to a ill-defined QRF associated to a superposed clock, hence showing that if the principle is violated there is a privileged reference frame, namely the one of the laboratory, which is inertial and not in superposition.

\section{Superposition of time dilations in atom interferometers}\label{sec:GTD}

\begin{figure}[t]
    \centering
    \includegraphics[width=\columnwidth]{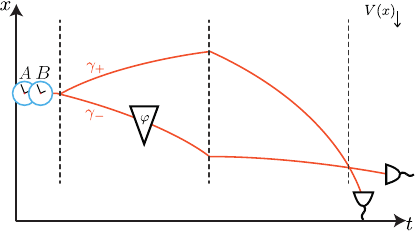}
    \caption{Two atomic clocks, A and B, in an interferometer vertically placed in a gravitational field $V(x)$. The setup is made of two beam splitters, a mirror, a controllable phase shifter $\varphi$, and two detectors. The upper trajectory is called $\gamma_+$ and the lower $\gamma_-$.}
    \label{fig:MZ}
\end{figure}

Atom interferometers have been used for precision tests of the classical equivalence principle (see e.g. Ref.~\cite{tino2020precision} for a review). An atom interferometer splits the trajectory of an atom into two paths in a quantum superposition. At the end of the interferometer, the paths are recombined and the relative phase shift contains information about the gravitational potential. However, when only the path degree of freedom is considered, the phase difference measured in atomic interferometers cannot be distinguished from the one originating from a suitably chosen non-gravitational potential. This means that all gravitational effects of these setups can be simulated or cancelled by non-gravitational ones. To overcome this problem, in Ref.~\cite{zych} it was shown that adding atomic clocks leads to a reduction of visibility in the interferometer, and that this effect can only be explained as coming from gravitational time dilation. This is a genuine interplay of general relativistic effects and quantum superposition: according to the principle of complementarity, the accumulation of a different proper time of the clock in the two paths due to gravity leads to an increase of distinguishability, and hence a reduction of visibility.

We consider the setup illustrated in Fig.~\ref{fig:MZ}, where two atomic clocks, $A$ and $B$, enter an interferometer in an entangled state from the perspective of the laboratory $L$. The interferometer is placed vertically in the gravitational field $\hat{V}(\hat{\mathbf x})$ of the Earth. Each clock acquires a different proper time according to the path it takes in the interferometer~\cite{zych, zychThesis}, resulting in a relative phase before the second beam splitter. Each clock $\X =\A , \B$ lives on a Hilbert space $\mathcal{H}_\X = \mathcal{H}_\Xe^{ext} \otimes \mathcal{H}_\Xi$, where $\mathcal{H}_\Xe^{ext}$ are the external degrees of freedom corresponding to the position or momentum of the clock, and $\mathcal{H}_{C_X} \simeq L^2 (\mathbb{R})$ is the internal degree of freedom of the clock, ticking according to its proper time. We assume an ideal clock, meaning that each instant of time is associated to an orthogonal state. This simplification can be thought as an idealisation of more complex and realistic models  that are physically realizable (see Ref.~\cite{Woods2018Autonomous} for details).

In the laboratory frame, we define the state of the clocks $\A$ and $\B$ at the time $\tau_L$ shown by the laboratory clock as $\ket{\psi^{(L)}(\tau_L)}_{\A \B}$. The Hamiltonian governing the time evolution of $\A$ and $\B$ is $\hat{H}^{(L)} = \sum_{I=A,B}( m_Ic^2 + \hat{H}_I^{(L)})$ (see Ref.\,\cite{zych} for details), where the energy shift due to the rest mass is added for future convenience, and 
\begin{equation}\label{eq:Hamiltonian_lab}
    \hat H_I^{(L)}=  \frac{\hat{\mathbf p}_I^2}{2m_I} + m_I V(\hat{\mathbf x}_I) + \hat H_{I}\left(1+\frac{V(\hat{\mathbf x}_I)}{c^2} -\frac{\mathbf{\hat p}_I^2}{2m_I^2c^2}\right).
\end{equation}
The operator multiplying the internal Hamiltonian of the clocks accounts for the first-order correction to proper time due to the general-relativistic and special-relativistic time dilation. It is then easy to see that, since time dilation is encoded in an operator depending on the position or momentum of the clock, when the position or momentum of the clock are in a quantum superposition state, its proper time also runs differently in each amplitude, i.e. it is in quantum superposition. This situation is more general than the typical scenario in which the classical EEP is formulated.

We now review how this setup can indeed be used to measure effects arising from a quantum superpositions of special-relativistic and gravitational time dilations: we show that the proper time of the clock is different in each amplitude due to velocity and gravity, and this yields a measurable phase difference.
For simplicity, we restrict to the regime in which the external degrees of freedom can be treated in the semiclassical limit, meaning that they appear only as fixed functions inside the Hamiltonian~\cite{zych, zychThesis}: their time evolution consists only in a phase. Under this approximation, we can simplify the evolution of the clocks in the interferometer by assigning them two states, e.g. $\{\ket{x_+}_\Ae,\ket{x_-}_\Ae\}$, corresponding to the upper and lower path $\gamma_\pm$ respectively. 

The initial state after the first beam splitter is (see Fig.~\ref{fig:MZ})
\begin{equation}\label{eq:initial_state_LZ}
    \ket{\psi_0^{(L)}}_{\A \B} = \frac{\ket{x_+}_\Ae\ket{x_+}_\Be + e^{2i\varphi}\ket{x_-}_\Ae \ket{x_-}_\Be}{\sqrt{2}}\ket{\tau_{in}}_\Ai \ket{\tau_{in}}_\Bi,
\end{equation}
where $\varphi$ is a phase that can be experimentally controlled, and from now on we choose $\tau_{in}=0$ for simplicity. The evolution inside the interferometer is different for each trajectory $\gamma_\pm$, and is given by the time evolution operator $\hat U_\pm = e^{-\frac{i}{\hbar} \int_{\gamma_\pm} dt \hat H^{(L)}}$. The state inside the interferometer, after the time evolution, is
\begin{align}\label{eq:state_semicl_one_particle}
    &\ket{\psi^{(L)} (\tau_L)}_{\A \B} = \frac{1}{\sqrt{2}}(\ket{x_+}_\Ae \ket{x_+}_\Be \ket{\tau_+}_\Ai \ket{\tau_+}_\Bi + \nonumber \\ 
    &\quad + e^{\mathrm{i}\left(2\varphi +\sum_{j=A,B}\Delta\phi_j\right)} \ket{x_-}_\Ae \ket{x_-}_\Be \ket{\tau_-}_\Ai \ket{\tau_-}_\Bi),
\end{align}
where $\Delta \phi_j = \frac{1}{\hbar} \int_{\Delta \gamma} dt \left(m_j c^2 + \frac{\mathbf{p}_j^2}{2m_j} + V(\mathbf{x}_j)\right)$ is the line integral along the closed path formed by the two arms of the interferometer, i.e. $\int_{\Delta \gamma}(\cdots) = \int_{\gamma_+}(\cdots) - \int_{\gamma_-}(\cdots) $ and $\ket{\tau_\pm}_{C_j} = e^{-\frac{i}{\hbar} \int_{\gamma_\pm} dt \left(1+\frac{V(\mathbf{x}_j)}{c^2}-\frac{\mathbf{p}_j^2}{2m_j^2c^2}\right) \hat H^j_{int}} \ket{\tau_{in}}_{C_j}$, $j=A,B$. This phase difference comes from the fact that the clocks $A$ and $B$ tick in a quantum superposition of different proper times, due to special-relativistic and general-relativistic time dilation.

We now show that this phase difference is measurable. The interferometric measurements are projections on the Hilbert space of the path alone (the clock is not measured)
\begin{equation} \label{eq:semicl_Leasurement}
    \ket{D_\pm}_{\Ae \Be} = \frac{\ket{x_+}_\Ae \ket{x_+}_\Be \pm \ket{x_-}_\Ae \ket{x_-}_\Be}{\sqrt{2}}.
\end{equation}
The measurement probabilities for two identical masses $m_A=m_B=m$ are $P_\pm = |\braket{D_\pm | \psi^{(L)}(\tau_L)}_{AB}|^2$, and result in
\begin{equation}\label{eq:prob_MZ}
    P_\pm = \frac{1}{2}\left(1\pm \left|\braket{\tau_+|\tau_-}\right|^2 \cos{(2\Delta \phi + 2\varphi + 2\varphi^\prime)}\right),
\end{equation}
where $\Delta \phi = \Delta \phi_A = \Delta \phi_B$ and $\varphi^\prime$ is defined by $\braket{\tau_+|\tau_-}_\Ai = \braket{\tau_+|\tau_-}_\Bi = \left|\braket{\tau_+|\tau_-}\right|e^{i\varphi^\prime}$.

The phase difference in Eq.~\eqref{eq:prob_MZ} encodes the dependence on the quantum superposition of special-relativistic and gravitational time dilations between the two interferometric paths. Classical EEP predicts special-relativistic and general-relativistic time dilation, while here we observe a quantum superposition thereof. We'll show that this is a prediction of the EEP for QRFs (see more in Section~\ref{sec:QRF_perspective}). This suggests that when the principle is violated, the probabilities of Eq.~\eqref{eq:prob_MZ} change in a measurable way, and this is indeed what we show in the next section. 

\section{Generalisation of the EEP to quantum particles and its violation}\label{sec:model}
In the previous section we showed that atomic clock interferometers with entangled clocks are sensitive to quantum superpositions of gravitational and special-relativistic time dilation, that are paradigmatic effects predicted by the EEP for QRFs. We now give a procedure to use atomic clock interferometers to test the EEP for QRFs. We introduce a model for violation of the principle and show that atomic clock interferometers can measure the violation parameters we introduce, hence providing a test for the principle.

The standard version of the EEP has been tested with experiments that prove the validity of each of the three aspects of the principle (WEP, LLI, LPI). For example, a paradigmatic test of LPI is the gravitational redshift experiment (see e.g. Ref.~\cite{will_book}), which measures the difference in frequency of two clocks at rest at two different heights $x_1 ,x_2$ in a gravitational field: one clock emits a signal, which is received by the other at a different frequency. To illustrate this experiment, we assume that WEP and LLI are valid, while LPI is not. The violation of LPI implies that in the freely falling frame that is momentarily at rest with respect to the clock, the proper time of the clock depends on the position of the clock, for example through the gravitational potential $\tau = \tau(V)$. We can describe this situation using some static coordinates $(t_s,x_s)$, that are accelerated with acceleration $g$ with respect to the previous freely falling frame. Let us assume for simplicity that the clocks are close, in such a way that the gradient of acceleration is negligible, namely $g = k \nabla \phi$ is constant between the clocks. The argument can be easily generalized to include also non-uniform accelerations. The relationship between the freely falling coordinates $(t_f,x_f)$, and the static ones is a sequence of Lorentz transformations. As a consequence, the gravitational redshift parameter is
\begin{equation}
    Z=\frac{\Delta \nu}{\nu} = 1 - \frac{\tau(V_{1})(1+g x_1 / c^2)}{\tau(V_{2})(1+g x_2 / c^2)}.
\end{equation}
Expanding $\tau(V) = \tau_0 + \frac{\partial \tau}{\partial V}\frac{g \Delta x}{k}$, we find 
\begin{equation}
    Z = \left(1 + \alpha \left(x \right)\right) \frac{\Delta V}{c^2},
\end{equation} where $\alpha(x) = -\frac{c^2}{k \tau_0} \frac{\partial \tau}{\partial V}$. We conclude that a violation of LPI implies a modification of the gravitational redshift factor, with a position-dependent parameter of violation $\alpha(x)$, which is zero if LPI is true. In particular, this means that if LPI is true, then two clocks at two different heights experience a time dilation factor $Z = \frac{\Delta V}{c^2}$, independently of the nature of the clocks: this is the universality of the gravitational redshift and the paradigmatic consequence of LPI.

We now show that this analysis extends naturally to Q-LPI. In particular, we now generalize the previous argument to QRFs associated to quantum clocks in a quantum superposition of positions in the gravitational field, namely we impose that the outcome of local experiments depend on the (quantum) position of the QRF where they are performed. If the clock is not localized, but it is in a quantum superposition, its position must be treated as an operator. Hence, instead of the parameter $\alpha(x)$, we should measure an operator $\alpha (\hat x)$. This means that there are more parameters to measure in the quantum version of the EEP: the out-of-diagonal elements of $\alpha(\hat x)$ are responsible for the quantum violation of the principle, while the diagonal elements are responsible for the classical violation. 

Similarly to the LPI violation parameter $\alpha(x)$, classical tests of the EEP involve the WEP violation parameter $\eta(x,p)$ and the LLI violation parameter $\beta(p)$. Here, we generalize them to $\eta(\hat x, \hat p)$ and $\beta(\hat p)$ following an analogous reasoning.

To introduce these violation operators, we modify the Hamiltonian of the systems. We follow the same strategy of Ref.~\cite{zych_QEEP}, where a model for violation of a different quantum  version of the EEP was introduced. The model in Ref.~\cite{zych_QEEP} uses the mass-energy equivalence to introduce mass operators: the energy of the internal degrees of freedom, namely the clock, is accounted for in the mass, and since it is quantised, also the mass is quantised. Ref.\,\cite{zych_QEEP} uses three different types of mass-operators $\hat{M}^{(s)} = m^{(s)} \mathbb{1} + \frac{\hat{H}^{(s)}}{c^2}$, with $s= g, i, r$, acting on the internal degrees of freedom of the clocks. In particular, the gravitational mass $\hat M^{(g)}$ couples to the gravitational field, the inertial mass $\hat M^{(i)}$ couples to the momentum, and the rest mass $\hat M^{(r)}$ is the mass in the rest frame of the atomic clock. The EEP, in its generalised form, is valid if all matrix elements of these operators are the same, namely if and only if $\hat M^{(g)} = \hat M^{(i)}$ (WEP), $\hat M^{(r)} = \hat M^{(i)}$ (LLI), and $\hat M^{(g)} = \hat M^{(r)}$ (LPI). 

The test of our generalisation of the EEP to QRFs requires a more general model for violation, because, differently to Ref.~\cite{zych_QEEP}, which regards only internal degrees of freedom, the quantum aspects of the EEP for QRFs arise when the external degrees of freedom of the clocks are in a quantum superposition state. Hence, we extend the model of violations of Ref.~\cite{zych_QEEP} by allowing an explicit dependence of position and momentum on the mass operator. For example, the most general violation of Q-LPI should also include an explicit position dependence on the violation operator, as shown before. An analogous conclusion can be drawn for Q-LLI for momentum-dependence. Therefore, we let   $\hat H^{(g)} = \hat H^{(g)}(\hat{\mathbf x}) = f(\hat A, \hat{\mathbf x})$, and $\hat H^{(i)} = \hat H^{(i)}(\hat{\mathbf p}) = g(\hat B, \hat{\mathbf p})$, namely $\hat H^{(g)}(\hat{\mathbf x})$ is a generic function of the position operator $\hat{\mathbf x}$ and of an operator $\hat A$ that acts on the internal Hilbert space only, and analogously for $\hat H^{(i)}(\hat{\mathbf p})$.

The violations of the three conditions composing the definition of the EEP are 
\begin{align}
    m^{(i)} \hat{\mathbb{1}}+ \hat H^{(i)}(\hat{\mathbf p}) &\neq m^{(g)} \hat{\mathbb{1}} + \hat H^{(g)} (\hat{\mathbf x}) & (\text{Q-WEP}), \label{eq:test_WEP_violation_condition}\\
    \hat H^{(r)} &\neq \hat H^{(i)} (\hat{\mathbf p}) & (\text{Q-LLI}), \label{eq:test_LLI_violation_condition}\\
    \hat H^{(r)} &\neq \hat H^{(g)} (\hat{\mathbf x}) & (\text{Q-LPI}), \label{eq:test_LPI_violation_condition}
\end{align}
where the rest mass does not appear in the second and third condition because it is not observable~\cite{zych_QEEP}. 
The conditions in Eqs.~(\ref{eq:test_WEP_violation_condition}~-~\ref{eq:test_LPI_violation_condition}) quantify the corresponding statements of the three aspects of the principle given in the introduction. The Hamiltonian of Eq.~\eqref{eq:Hamiltonian_lab} is modified to
\begin{equation}\label{eq:test_violated_hamiltonian}
   \begin{split}
        \hat H_{I,v}^{(L)} & = \frac{\hat{\mathbf p}_I^2}{2m_I^{(i)}} + m_I^{(g)} V(\hat{\mathbf x}_I) + \hat H_I^{(r)} +\\
    &+\hat H_I^{(g)} (\hat{\mathbf x}_I) \, \frac{V(\hat{\mathbf x}_I)}{c^2} - \hat H_I^{(i)} (\hat{\mathbf p}_I) \, \frac{\hat{\mathbf p}_I^2}{2m_I^{(i)2}c^2}.
   \end{split}
\end{equation}
Note that the dependence on the phase-space operators does not introduce any ordering problem.

We now define the violation operators described above as
\begin{align}
    \hat \eta (\hat{\mathbf x}, \hat{\mathbf p}) &= \mathbb{\hat 1} - \hat M^{(g)} (\hat{\mathbf x})\, \hat M^{(i)-1}(\hat{\mathbf p}) & (\text{Q-WEP}), \label{eq:model_eta_new}\\
    \hat \beta (\hat{\mathbf p})&= \mathbb{\hat 1} - \hat H^{(i)} (\hat{\mathbf p})\, \hat H^{(r)-1} & (\text{Q-LLI}), \label{eq:model_beta_new}\\
        \hat \alpha (\hat{\mathbf x}) &= \mathbb{\hat 1} - \hat H^{(g)} (\hat{\mathbf x})\, \hat H^{(r)-1} & (\text{Q-LPI}),\label{eq:model_alpha_new}
\end{align}
with $\hat{\eta}$, $\hat{\beta}$, and $\hat{\alpha}$ being invertible to our order of approximation. 

We now show that the interferometric setup described previously is sensitive to such violation of the EEP for QRFs, and consequently can provide a test for it.

\subsection{Violation of Q-LPI}\label{sec:test_LZ_LPI}
For a violation of Q-LPI, we impose the condition $\hat{H}^{(r)} \neq \hat{H}^{(g)}(\hat{\mathbf{x}})$ in Eq.~\eqref{eq:test_LPI_violation_condition}. In the laboratory frame $L$, we obtain the Hamiltonian $\hat{H}^{(L)}_{LPI} = \sum_{j= A, B}\left( m_j c^2+ \hat{H}^{LPI}_{j}\right)$, with
\begin{equation}\label{eq:test_H_LZ_LPI}
    \begin{split}
    \hat{H}^{LPI}_{j} &= \frac{\mathbf{p}_j^2}{2m_j} + m_j V(\mathbf{x}_j) + \hat H_j^{(r)} + \nonumber \\  &+\hat H_j^{(g)} (\mathbf{x}_j) \frac{V(\mathbf{x}_j)}{c^2} - \hat H_j^{(r)} \frac{\mathbf{p}_j^2}{2m_j^2c^2}.
    \end{split}
    \end{equation}
In the interferometric setup we have considered, the probabilities are formally the same as in Eq.~\eqref{eq:prob_MZ}, but the factor $\braket{\tau_+ | \tau_-}^\prime_j$, $j=A,B$, is different. To the lowest order in the violation operator $\hat \alpha(\mathbf{x})$, discarding the commutators between $\hat \alpha(\mathbf{x})$ and all the other operators, we find $\Delta_{LPI} \braket{\tau}_j = \braket{\tau_+ | \tau_-}^\prime_j - \braket{\tau_+ | \tau_-}_j$, with
\begin{equation}
    \Delta_{LPI} \braket{\tau}_j= - \frac{i}{\hbar}\int_{\Delta \gamma} dt \ \frac{V(\mathbf{x}_\gamma)}{c^2}\alpha(\mathbf{x}_\gamma)\,{}_j \!\bra{\tau_+} \hat H^{(r)} \ket{\tau_-}_j,
\end{equation}
where $\Delta \gamma$ is the closed path formed by the two arms of the interferometer and $\mathbf{x}_\gamma$ is the position evaluated along the (classical) trajectory of the system. Consequently, the probability explicitly depends on the matrix elements of the violation operator, which means that the experiment is sensitive to violations of Q-LPI.

\subsection{Violation of Q-LLI}
For a violation of Q-LLI, we impose the condition $\hat{H}^{(r)} \neq \hat{H}^{(i)}(\hat{\mathbf{p}})$ in Eq.~\eqref{eq:test_LLI_violation_condition}. With analogous notation to the case of Q-LPI, we obtain 
\begin{equation}
   \Delta_{LLI} \braket{\tau}_j = \frac{i}{\hbar}\int_{\Delta \gamma} dt \ \frac{\mathbf{p}_\gamma^2}{m^2c^2}\beta(\mathbf{p}_\gamma) \,{}_j \!\bra{\tau_+} \hat H^{(r)} \ket{\tau_-}_j,
\end{equation}
Thus, the experiment is sensitive to violations of LLI.

\subsection{Violation of Q-WEP}
The violation of Q-WEP is encoded in the operator    $\eta (\mathbf{\hat x},\mathbf{\hat p}) = \mathbb{\hat 1} - \hat M^{(g)}(\mathbf{\hat x}) \, \hat M^{(i)-1}(\mathbf{\hat p})$, where $\hat M^{(g)} (\mathbf{\hat x}) = m^{(g)} + (1-\hat \alpha (\mathbf{\hat x})) \hat H^{(r)}$ and $\hat M^{(i)} (\mathbf{\hat p}) = m^{(i)} + (1- \hat \beta (\mathbf{\hat p})) \hat H^{(r)}$. Moreover, the standard WEP violation is $ \eta = 1 - \frac{m^{(g)}}{m^{(i)}}$. Hence $\eta (\hat x,\hat p)$ is a function $\eta = \eta \left(\alpha(\mathbf{\hat x}), \beta(\mathbf{\hat p}), \eta, m^{(i)}, \hat H^{(r)} \right)$. Provided that the inertial mass $m^{(i)}$ and the rest Hamiltonian $\hat H^{(r)}$ are known, $\hat \eta$ is fully determined by $ \alpha(\mathbf{\hat x})$, $\beta(\mathbf{\hat p})$ and the classical parameter $\eta$. Since we already showed that the Mach-Zehnder interferometer is sensitive to the coefficients of $\alpha(\mathbf{\hat x})$ and $\beta(\mathbf{\hat p})$, it only remains to prove its sensitivity to $\eta$, namely that it can provide a classical standard WEP test.

Therefore, by imposing the condition for the mass of particle $j=A,B$ $m_j^{(g)} \neq m_j^{(i)}$ and by neglecting the rest-mass term, we find that the Hamiltonian in the laboratory frame in Eq.~\eqref{eq:Hamiltonian_lab} is modified to $\hat{H}^{(L)}_{WEP} = \sum_{j= A, B}\hat{H}^{WEP}_{j}$, where
\begin{equation}
   \hat{H}^{WEP}_{j} = \frac{\mathbf{p}_j^2}{2m_j^{(i)}} + m_j^{(g)} V(\mathbf{x}_j) + \hat H_j \left(1 + \frac{V(\mathbf{x}_j)}{c^2} - \frac{\mathbf{p}_j^2}{2m^{(i)2}_j c^2}\right).
\end{equation}
The measurement probabilities are the same of Eq.~\eqref{eq:prob_MZ}, where
\begin{equation}
    \Delta \phi_j = \frac{1}{\hbar}\int_{\Delta \gamma} dt \ \left[ (1-\eta)m^{(i)}_j V(\mathbf{x}_j) + \frac{\mathbf{p}_j^2}{2m_j^{(i)}}\right].
\end{equation}
Thus, the setup can provide a test for WEP. We have thus shown that the atomic clock interferometer can test all of the three aspects of the EEP for QRFs.

\section{The EEP for QRFs in the perspective of a quantum particle}\label{sec:QRF_perspective}

We now use our interferometetric setup to illustrate the three aspects of the EEP for QRFs, presented in Section~\ref{sec:model}, and show what the predictions of the principle for the proper time of a superposed clock are. As explained in the previous sections, the laboratory clock and the clocks in the interferometer are in a quantum relation relative to each other. These are not related via a standard reference frame transformation, but via a QRF transformation. In particular, we need to transform all the spacetime coordinates (in our case, we are in $1+1$D), and write the Schr{\"o}dinger equation in the two frames in the proper time of the two clocks. To achieve this goal, the most suitable formulation is the one introduced in Ref.\,\cite{giacomini_SQRF}, which we briefly sketch here and comprehensively review in Appendix~\ref{sec:SQRF}.

The SQRFs formalism is a generalisation of the QRFs formalism~\cite{giacomini_QRF} that additionally treats space and time on an equal footing and gives a timeless and fully relational description of a set of physical systems from the point of view of one of them. It adopts elements of Covariant Quantum Mechanics~\cite{reisenberger} and the Page-Wootters mechanism~\cite{page, giovannetti}. Moreover, it accounts for both external and internal degrees of freedom of quantum systems, that are both employed in building the SQRF description. 

The SQRFs formalism is a timeless formulation of a set of quantum systems, where the dynamical evolution of the systems is encoded in a set of constraints and emerges through a procedure which fixes the redundancies induced by the constraints. Such a procedure has the physical interpretation of a reduction to the QRF of one of the quantum systems considered. Each system has an external Hilbert space, corresponding to the position or momentum of the centre of mass of the system, and an internal Hilbert space, corresponding to a clock. Both external and internal degrees of freedom are used to identify the QRF: the external ones are used to fix the transformation to the QRF, and the internal ones identify the proper time in the QRF of each system. The formalism is completely relational, meaning that there is no external spacetime structure beside the relations between the particles, as an effect of imposing the constraints.

The physical state of the atomic clocks and the laboratory satisfies the constraint $\hat C \ket{\Psi}_\text{ph} = 0$, where $\hat C$ is a linear combination of different constraints, namely the dispersion relation $\hat C_I$ for each particle $I$, the conservation of total energy $\hat f^0$ and the conservation of total momentum $\hat f^1$ (details in Appendix~\ref{sec:SQRF}).

The perspective of a particle is obtained by applying a unitary operator $\mathcal{\hat T}_i$ to the physical state, to transform the coordinates to the relative coordinates with respect to particle $i$, and set the metric to be flat at the position of the particle. Finally, the resulting state is projected on a state to fix the origin of the reference frame to be in the position of the particle. For example, for the laboratory:
\begin{equation}\label{eq:QRF_state_particle_1}
    \ket{\psi}^{(L)} = \bra{q_L=0} \mathcal{\hat T}_L \ket{\Psi}_{ph},
\end{equation}
where the specific form of $\mathcal{\hat T}_L$ is given in Appendix~\ref{sec:SQRF}. Through a lengthy but straightforward calculation, one finds
\begin{equation} \label{eq:historystate_L}
    \ket{\psi}^{(L)} \propto \int d\tau_L \ket{\psi^{(L)}(\tau_L)}\ket{\tau_L}_\Li ,
\end{equation}
where $\ket{\psi^{(L)}(\tau_L)}_{\A \B}=e^{-\frac{\mathrm{i}}{\hbar} \hat{H}^{(L)}\tau_L}\ket{\psi_0^{(L)}}_{\A \B}$, $\ket{\psi_0^{(L)}}_{\A \B}$ is the initial state of the systems in the full Hilbert space.

This state is a so-called ``history state'', usually found in timeless approaches to quantum theory. Given a quantum system at time $\tau_L$, its quantum state can be described as $\ket{\psi^{(L)}(\tau_L)}$. When the clock is a quantum degree of freedom, we can write all possible states of the system at different times as being correlated, obtaining the ``history state'' in Eq.~\eqref{eq:historystate_L}.

The usual state at an arbitrary time $t$ in the perspective of the laboratory can be retrieved with a projection on the state of the clock, namely    $\ket{\psi^{(L)}(t)}_{\A \B} = {}_{C_L}\braket{t | \psi}^{(L)} = e^{-\frac{i}{\hbar}\hat H^{(L)} t} \ket{\psi_0^{(L)}}_{\A \B}$, where $\hat H^{(L)}$ is the Hamiltonian in Eq.~\eqref{eq:Hamiltonian_lab}. Hence the predictions of the SQRF formalism coincide with standard quantum mechanics in the perspective of the laboratory.

If also the perspective of the clocks in the interferometer is valid to describe the evolution of external systems to them, an analogous ``history state'' to Eq.~\eqref{eq:historystate_L} should describe the dynamical evolution in the perspective of one of the clocks, say clock $A$. The SQRF formulation provides a method to achieve this transformation, which cannot be obtained with other formulations of QRFs. The ``history state'' of $B$ and $L$ after the QRF transformation from L to A (see again Appendix~\ref{sec:SQRF} for details) is
\begin{equation}
    \ket{\psi}^{(A)} \sim \int d\tau_A \ket{\psi^{(A)}(\tau_A)}_{\B \L} \ket{\tau_A}_\Ai,
\end{equation}
where $\ket{\psi^{(A)}(\tau_A)}_{\B \L} = e^{-\frac{\mathrm{i}}{\hbar}\hat{H}^{(A)}\tau_A}\ket{\psi^{(A)}_0}_{\B \L}$ and the notation is completely analogous to the one in Eq.~\eqref{eq:historystate_L}. The Hamiltonian (see Appendix~\ref{sec:SQRF} for details) is
\begin{align}
    &\hat{H}^{(A)} = \sum_{j=B,L} \hat{T}_{j} + \sum_{J=A,B,L}\left(m_J c^2 + \hat{T}^\prime_J\right) + m_B \hat{V}'(\hat{\mathbf{q}}_B, \hat{\mathbf{q}}_L) + \nonumber\\ 
    & - m_L V(\hat{\mathbf q}_L) +\hat{\Delta}^{{(A)}}_{B}\hat H_B + \hat{\Delta}^{(A)}_L \hat H_L,
\end{align}
where we have defined $\hat{T}_{j} = \frac{\hat{\mathbf k}_j^2}{2m_j}$ for $j=B,L$ and $\hat{T}'_{J}=\frac{(\sum_{j=B,L} \hat{\mathbf k}_j)^2}{2m_A^2}m_J$ for $J=A,B,L$, and in addition $\hat{V}'(\hat{\mathbf{q}}_B, \hat{\mathbf{q}}_L)=V(\hat{\mathbf q}_B-\hat{\mathbf q}_L) - V(\hat{\mathbf q}_L)$, $\hat{\Delta}^{{(A)}}_{B} = \left(1+\frac{\hat{V}'(\hat{\mathbf{q}}_B, \hat{\mathbf{q}}_L)}{c^2} + \frac{\hat{T}'_A}{m_A c^2} - \frac{\hat{T}_B}{m_B c^2}\right)$, $\hat \Delta^{(A)}_L = \left(1-\frac{V(\hat{\mathbf q}_L)}{c^2} + \frac{T'_A}{m_Ac^2}- \frac{\hat T_L}{m_L c^2}\right)$. This Hamiltonian is consistent with the one obtained in the  literature of QRFs, and reduces to the standard Hamiltonian of quantum particles interacting via a potential in an appropriate classical limit of the QRFs (see Ref.~\cite{giacomini_SQRF} for details).

This Hamiltonian encodes the three aspects of the EEP for QRFs: it shows the evolution of the clock $B$ with respect to the proper time of the clock $A$, when they are in quantum superposition with respect to each other. For simplicity in the following we only illustrate the consequence of Q-LPI, neglecting special-relativistic time dilation, i.e. assuming that in the two paths of the interferometer the velocity is the same, or the difference is so small it can be neglected.

The quantum state in the laboratory frame in Eq.~\eqref{eq:state_semicl_one_particle} is transformed to the perspective of $A$ as $\ket{\psi^{(A)}(\tau_A)}_{\B \L} =  \frac{1}{\sqrt{2}}\left( \ket{\Psi(\tau_A,x_+,x_+)}_{\Be \L} + \ket{\Psi(\tau_A,x_-,x_-)}_{\Be \L}\right) \ket{\tau_A}_\Bi$, where the explicit expression of $\ket{\Psi(\tau_A,x_\pm, x_\pm)}_{\Be \L}$ is given in Appendix~\ref{sec:appa_QUGR}. This shows that, even if the reading of time of $C_A$ and $C_B$ was unsharp in the laboratory frame, due to the quantum superposition, in the perspective of one of the clocks the time reading of the other clock is well-defined and there is no time dilation. This is an extension of the \emph{universality of gravitational redshift} to quantum superpositions, which is a consequence of Q-LPI, analogously to the classical universality of gravitational redshift being a consequence of LPI, as detailed in Section~\ref{sec:model}.

Similarly, if we had started from an initial state of the form
\begin{equation}
    \ket{\psi_0^{(L)}}_{\A \B} = \frac{\ket{x_+}_\Ae \ket{x_-}_\Be + \ket{x_-}_\Ae \ket{x_+}_\Be}{\sqrt{2}}\ket{\tau_{in}}_\Ai\ket{\tau_{in}}_\Bi,
\end{equation}
where in each amplitude the two clocks were at different heights, we would have found the state in the perspective of the clock $A$ to be (see Appendix~\ref{sec:appa_QUGR} for details): 
\begin{align}
    \ket{\psi^{(A)}(\tau_A)}_{\B \L} = \frac{1}{\sqrt{2}}\big(&\ket{\Psi(\tau_A,x_+,x_-)}_{\Be \L}\ket{\tau_A + \Delta \tau}_\Bi + \nonumber \\&
    \ket{\Psi(\tau_A,x_-,x_+)}_{\Be \L} \ket{\tau_A - \Delta \tau}_\Bi\big)
\end{align}
where $\Delta \tau = \frac{V(\mathbf{x}_+) - V(\mathbf{x}_-)}{c^2}\tau_A$.

We conclude that if two clocks are in spatial superposition, in each amplitude there is a time dilation of the second clock according to the first one, if in that amplitude the clocks are not at the same height.
Moreover, the dilation factor of each amplitude coincides with the classical dilation factor $\frac{\Delta \nu}{\nu} = \frac{\Delta V}{c^2}$. This is a genuine prediction of Q-LPI, analogously to the classical time dilation factor predicted by classical LPI. When the clocks are not in superposition, this form of quantum universality of gravitational time dilation reduces to the classical one, as exemplified in Appendix~\ref{sec:appa_QUGR}. A similar procedure leads to the generalization of special relativistic time dilation to quantum superpositions.

\section{Violation of the EEP for QRFs in the perspective of a quantum particle}\label{sec:test_LZ}
In the previous section we showed what the predictions of the EEP for QRFs are in the perspective of a clock in a superposed state.  We now show that the violation of the EEP for QRFs introduced in Section~\ref{sec:model} would lead to the dependence of the proper time of the clock on its position or momentum operator, thus generalising the classical violation. Detecting such a violation experimentally would imply that the EEP cannot be extended to QRFs, at least in the proposed form, and hence challenge the equivalence between all QRFs, highlighting the existence of a preferred QRF, coinciding in this case with the laboratory frame.

To obtain the modified Hamiltonian encoding the violations of the EEP for QRFs, we modify the total energy constraint of the SQRF formalism (see Appendix~\ref{sec:SQRF} for a detailed description), which contains the dependence on the internal Hamiltonians.
If the EEP for QRFs is violated, the total energy constraint $\hat{f}^0$ of Eq.~\eqref{eq:constraint_f0} is modified to
\begin{align}
    \hat f^0 &= \sum_{I=A,B} \hat p_I^0 + \left(\sqrt{g_{00}(\hat{\mathbf x}_I - \hat{\mathbf x}_L)}-1\right)\frac{\hat H_I^{(g)}}{c} + \nonumber \\
    &+\left(\sqrt{1+\frac{\hat{\mathbf p}_I^2}{m_I^2 c^2}}-1\right)\frac{\hat H_I^{(i)}}{c} + \frac{\hat H_I^{(r)}}{c} + \hat p_L^0.
\end{align}
We assume that, even in the presence of a violation, it is always possible to take the perspective of the laboratory, in which the dynamics follows a standard Hamiltonian evolution. Without this assumption, it would be impossible to even devise a test of the generalised EEP. In addition, the validity of the laboratory perspective naturally follows from the construction in Appendix~\ref{sec:SQRF}.

If any aspect of the EEP for QRFs (Q-LPI, Q-LLI, Q-WEP) is violated, as in Eqs.\,(\ref{eq:test_WEP_violation_condition}-\ref{eq:test_LPI_violation_condition}), it is impossible to find a ``history state'' in the perspective of $A$ or $B$. We illustrate this fact for Q-LPI, but the reasoning is the same for Q-LLI and Q-WEP. The explicit calculations for all cases are detailed in Appendix~\ref{sec:appb}. 

Specifically, let us assume that $\hat{H}_j^{(r)} \neq \hat{H}_j^{(g)}(\hat{\mathbf{x}}_j)$ but $\hat{H}_j^{(r)} = \hat{H}_j^{(i)}(\hat{\mathbf{p}}_j)$ for $j=A,B$. When we reduce to the perspective of one of the systems, say $A$, we find a quantum state of the form
\begin{equation}
    \ket{\psi}^{(A)} \sim \int d\tau_A dq_L dt_A \ket{\chi(\tau_A,\mathbf{q}_L,t_A)}_{\B \L} \ket{\varphi(\tau_A,\mathbf{q}_L,t_A)}_\Ai,
\end{equation}
where the relevant aspect of the expression is that the quantum state of the clock whose proper time is used to describe the dynamical evolution explicitly depends on the coordinate $\mathbf{q}_L$ through the function $\varphi$. The explicit form of $\chi (\tau_A,\mathbf{q}_L,t_A)$ and $\varphi (\tau_A,\mathbf{q}_L,t_A)$ is not relevant to our argument, and is reported in Appendix~\ref{sec:appb}. The crucial point in our discussion is that the dependence on the position $\mathbf{q}_L$ and on the time $t_A$ in the previous expression is the quantum generalisation of the violation of UGR that we reviewed in Section~\ref{sec:model}. This result means that, in the QRF of system A, the clock behaves differently according to where it is placed in the gravitational field. This is not compatible with Q-LPI, and hence shows that the EEP for QRFs is violated.

We have seen that the violation of the EEP for QRFs prevents a description from the point of view of the clocks in superposition, since their QRF is not well-defined. Nevertheless, since the laboratory is localized, we can still give predictions from the point of view of the laboratory, and we have shown in Section~\ref{sec:model} that these predictions depend on the violation parameters: this means that the experiment described can be used to test the principle.

\section{Conclusion}\label{sec:conclusion}
In conclusion, we showed that the usual definition of the EEP is not sufficient for quantum test particles in a quantum superposition of positions and velocities. We proposed to generalise the three aspects that are tested in a classical test of the EEP, the weak equivalence principle, the local position invariance, and the local Lorentz invariance, to account for particles being in a quantum state. We then formulated a model to test the violation of such generalized EEP, and showed that an interferometric setup can be used to test this principle.
In addition, we showed that violations of the EEP for QRFs breaks the equivalence between the laboratory perspective and the perspective of the QRF associated to the clock in the interferometer, and that this entails the impossibility of describing the dynamical evolution of the physical systems external to the QRF from the point of view of one of the quantum particles in the interferometer. This implies the existence of a preferred reference frame, the laboratory frame, that is classical. Hence, confirming the EEP for QRFs is also a necessary condition for the validity of the QRF perspective. On the other hand, the verification of the validity of the QRF perspective introduces a new ``quantum relativity principle'', supporting the idea that the gravitational field acquires quantum properties~\cite{Overstreet:2022zgq}. 

Note that the validity of this principle can be verified by measuring outcome probabilities, and hence does not rely on any specific interpretation of quantum mechanics, as long as it gives the same experimental predictions as ordinary quantum mechanics.

A verification of this generalised principle would show that quantum theory and general relativity are more compatible than what is currently believed. This result could then be used as a guiding principle to gain a deeper insight into the interface between the two theories in more complex scenarios.

\acknowledgments{The authors would like to thank Matteo Paris for helpful discussions. This research was funded in whole or in part by the Austrian Science Fund (FWF) [10.55776/F71]. For open access purposes, the author has applied a CC BY public copyright license to any author accepted manuscript version arising from this submission. F.G. acknowledges support from Perimeter Institute for Theoretical Physics. Research at Perimeter Institute is supported in part by the Government of Canada through the Department of Innovation, Science and Economic Development and by the Province of Ontario through the Ministry of Colleges and Universities. F.G. acknowledges support from the Swiss National Science Foundation via the Ambizione Grant PZ00P2-208885, from the ETH Zurich Quantum Center, and from the John Templeton Foundation, as part of the \href{https://www.templeton.org/grant/the-quantuminformation-structure-ofspacetime-qiss-second-phase}{‘The Quantum Information Structure of Spacetime, Second Phase (QISS 2)’ Project}. The opinions expressed in this publication are those of the authors and do not necessarily reflect the views of the John Templeton Foundation.}

\bibliography{bibliography}

\appendix
\onecolumngrid

\section{Spacetime Quantum Reference Frames}\label{sec:SQRF}
In this section we review the Spacetime Quantum Reference Frames (SQRFs) formalism, introduced in Ref.~\cite{giacomini_SQRF}.  This formalism allows us to describe the reference frame associated to quantum particles that evolve in a gravitational field generated by another quantum particle, in the limit of weak gravitational fields and slow velocities. We will use this tool to highlight how the EEP for QRFs can be cast in terms of proper times of quantum clocks. 

The SQRFs formalism is a generalisation of the QRFs formalism~\cite{giacomini_QRF} that additionally treats space and time on an equal footing and gives a timeless and fully relational description of a set of physical systems from the point of view of one of them. It adopts elements of Covariant Quantum Mechanics~\cite{reisenberger} and the Page-Wootters mechanism~\cite{page, giovannetti}. Moreover, it accounts for both external and internal degrees of freedom of quantum systems, that are both employed in building the SQRF description. 

The SQRFs formalism is a timeless description of quantum systems, where their dynamical evolution is encoded in a set of constraints and emerges through a procedure that eliminates redundancies induced by these constraints. This procedure has the physical interpretation of reducing to the QRF of one of the quantum systems considered. Each system is described with an external Hilbert space, corresponding to its position or momentum, and an internal Hilbert space, corresponding to a clock. Both degrees of freedom are employed to identify the QRF associated to each system: the external ones fix the transformation to the QRF, while the internal ones determine the proper time in the QRF of each system. Moreover, the formalism is entirely relational: there is no external spacetime structure other than the relations between the particles.

\begin{figure}[t]
    \centering
    \includegraphics{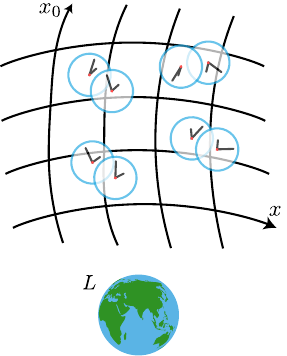}
    \caption{A system of $N$ non-interacting quantum particles in the gravitational field generated by a mass $L$. Each particle has both external and internal degrees of freedom, the latter acting as clocks. Therefore, the particles can be in a superposition of different positions, each associated to a different gravitational time dilation.}
    \label{fig:QRF_SQRF}
\end{figure}

Concretely, we consider a system of $N$ non-interacting quantum particles of mass $m_I$, $I=1,\dots,N$ in the weak gravitational field generated by a particle $L$ of mass $m_L$, as depicted in Figure~\ref{fig:QRF_SQRF}. Each particle lives on a Hilbert space $\mathcal{H}_\J = {\mathcal{H}_\Je}^{ext} \otimes \mathcal{H}_\Ji$, $J=1,\dots,N,L$ where ${\mathcal{H}_\Je}^{ext} \simeq L^2 (\mathbb{R}^2)$ corresponds to the position or momentum of the system in $(1+1)$D (one time coordinate and one space coordinate), and $\mathcal{H}_\Ji \simeq L^2 (\mathbb{R})$ to the internal state of the clock, ticking according to its proper time. The physical state of the $N$-particle system and the source mass $L$ satisfies the constraint
\begin{equation}\label{eq:physical_state}
    \hat C \ket{\Psi}_{ph} = 0,
\end{equation}
where $\hat C$ is a linear combination of first-class constraints~\cite{dirac1950}, namely constraints that commute with each other. Specifically,
\begin{equation}\label{eq:QRF_old_total_C}
    \hat C = \sum_{J=1}^{L} \mathcal{N}_J \hat C_J +  z_\mu \hat f^\mu,
\end{equation}
where the spacetime labels $\mu = 0,1$ refer to a $(1+1)$-dimensional spacetime. In the following, the spatial component is written in boldface, while the 2-vector is in plain text, e.g. $x^\mu = \{x^0,\mathbf{x}\}$. The constraints describe both the dynamics of the particles and the global symmetries of the model, namely spacetime global translations: the symmetry constraints $\hat f^\mu$ enforce that total momentum and total energy are null, i.e., that the model is globally translational-invariant in space and time. This condition corresponds to having a model where the dynamics is relational~\cite{mercati2018shape, barbour2014identification, vanrietvelde}. The constraints $\hat C_J$ enforce the general-relativistic dispersion relation of each particle.

We consider a physical scenario where the the gravitational field sourced by the mass $L$ is weak, hence it can be described in the weak-field limit in terms of the metric
\begin{align}\label{eq:QRF_newtonian_Letric}
    g_{00} &= 1 + 2\frac{V(\mathbf{x}-\mathbf{x_L})}{c^2}, \\
    g_{01} &= g_{10} = 0, \\
    g_{11} &= -1,
\end{align}
which reproduces the effect of Newtonian gravity, but also allows for general-relativistic effects such as time dilation. The constraints in this regime are
\begin{align}
\hat C_I &= \sqrt{g^{00}(\hat{\mathbf{ x}}_I - \hat{\mathbf{x}}_L)} \hat p_0^I - \omega_\mathbf{p}^I, \label{eq:constraint_CI} \\
\hat C_L &= \hat p_0^L - m_L c - \frac{\hat{\mathbf p}_L^2}{2m_L c}, \label{eq:constraint_CL}\\
\hat f^0 &= \sum_{I=1}^N \left[\hat p_0^I + \Delta(\hat {\mathbf x}_I - \hat {\mathbf x}_L, \hat{\mathbf p}_I) \frac{\hat H_I}{c}\right] + \hat p_0^L + \frac{\hat H_L}{c}, \label{eq:constraint_f0}\\
    \hat f^1 &= \sum_{I=1}^N \hat{\mathbf p}_I + \hat{\mathbf p}_L. \label{eq:constraint_f1}
\end{align}
The operators $\hat x_I^\mu$ and $\hat p_\nu^I$ satisfy $[\hat x_I^\mu,\hat p_\nu^I] = i\hbar \, \delta^\mu_\nu$, for $I=1,\dots,N$, where $\hat x_I^0$ is the coordinate time operator. $\hat H_I$ is the internal Hamiltonian for each particle, associated to the internal evolution of the clock. Moreover, $\hat \omega_{\mathbf p}^I = m_I c \ \hat \gamma_I$ where $\hat \gamma_I = \sqrt{1+\frac{\hat{\mathbf p}_I^2}{m_I^2c^2}}$. Here $\Delta(\hat {\mathbf x}_I - \hat {\mathbf x}_L, \hat{\mathbf p}_I)$, with $I=1, \cdots, N$ is the worldline operator of a quantum relativistic particle in a weak gravitational field, and is responsible for the time dilation of the quantum clock, as explained in detail in Ref. \cite{giacomini_SQRF}, and it reads $\Delta(\hat {\mathbf x}_I - \hat {\mathbf x}_L, \hat{\mathbf p}_I) = \sqrt{g_{00}(\hat{\mathbf x}_I - \hat {\mathbf x}_L)}\, \hat \gamma_I^{-1}$. All calculations are performed keeping only the leading terms in $O\left(\frac{V}{c^2}\right)$, $O\left(\frac{p}{mc}\right)$, and discarding factors $O\left(\frac{p^2 V}{m^2c^4}\right)$.

In addition to the original formulation given in Ref.~\cite{giacomini_SQRF}, we here additionally consider the dynamics of the mass $L$ that generates the gravitational field, through the dynamical constraint $\hat{C}_L$. 
Since particle $L$ is the only source of the gravitational field, its dynamical evolution is free and in Minkowski spacetime. Equivalently, this can be seen as a choice of local coordinates for all systems involved in the description, which we assume localised in a region of space (in our specific setup, in the laboratory in which the experiment is performed).

In our setting, $L$ represents the Earth, that is responsible for sourcing the gravitational field. We further consider a laboratory, whose position is constrained to the Earth: no relative motion exists of the laboratory relative to the Earth, hence we treat the two as a single quantum system. For this reasons, we refer to the reference frame of $L$ as the laboratory frame. The introduction of the dynamics for the mass $L$ is useful to give a description of the system from the point of view of the laboratory, and to see what a specific state in the perspective of the laboratory looks like in the perspective of another particle.

We consider the mass $m_L$ to be the mass of the Earth, meaning that $m_L \gg m_J$, with $J=1, \cdots, N$. As a consequence, we will find that the frame of $L$ is inertial in the Newtonian sense, reproducing the results of the dynamics of a quantum clock in a gravitational field, described by an inertial frame, as in Ref.~\cite{zych}.

The physical state can be obtained from Eq. \eqref{eq:physical_state} to be
\begin{equation}
    \ket{\Psi}_{ph} \propto \int d^{N+1}\mathcal{N}d^2 z \ e^{\frac{i}{\hbar}\mathcal{N}_J \hat C_J} e^{\frac{i}{\hbar} z_\mu \hat f^\mu} \ket{\phi},
\end{equation}
where $\ket{\phi}$ can be written as
\begin{equation}
    \ket{\phi} = \int \Pi_I\left[d\mu (x_I) dE_I\right] d^2x_L dE_L \phi(x_1,\dots,x_L,E_1,\dots,E_L) \ket{x_1,\dots,x_L} \ket{E_1,\dots,E_L},
\end{equation}
and $d\mu(x_I) = \sqrt{g_{00}(\mathbf{x}_I - \mathbf{x}_L)}d^2x_I$ is the covariant integration measure.

The perspective of a particle is obtained by applying a unitary operator $\mathcal{\hat T}_i$ to the physical state, to transform the coordinates to the relative coordinates with respect to particle $i$, and set the metric to be flat at the position of the particle. Finally, the resulting state is projected on a state to fix the origin of the reference frame to be in the position of the particle. For example, for particle 1:
\begin{equation}\label{eq:QRF_state_particle_1}
    \ket{\psi}^{(1)} = \bra{q_1=0} \mathcal{\hat T}_1 \ket{\Psi}_{ph},
\end{equation}
where $\mathcal{\hat T}_1$ is 
\begin{equation}\label{eq:QRF_transformation_operator_1}
    \mathcal{\hat T}_1 = e^{-\frac{i}{\hbar} \frac{\log{\sqrt{g_{00}(\hat {\mathbf x}_L)}}}{2}\sum_{J=1}^L \left(\hat x_J^0 \hat p_J^0 + \hat p_J^0 \hat x_J^0\right)}e^{\frac{i}{\hbar} \hat {\mathbf x}_1 (\hat f^1 - \hat {\mathbf p}_1)}e^{\frac{i}{\hbar} \hat x_1^0(\hat f^0 - \hat p_1^0)}.
\end{equation}

Through a lengthy but straightforward calculation, detailed in Ref.~\cite{giacomini_SQRF}, one can show that the state in the perspective of particle 1 is equal to a ``history state''~\cite{giovannetti, castro-ruiz, giacomini_SQRF}: an entangled state associating to each time state $\ket{\tau_1}_{C_1}$ of the internal clock of the particle a state $\ket{\psi^{(1)}(\tau_1)}$ describing all the other particles at time $\tau_1$ as 
\begin{equation} \label{eq:historystate_1}
    \ket{\psi}^{(1)} \propto \int d\tau_1 \ket{\psi^{(1)}(\tau_1)}\ket{\tau_1},
\end{equation}
where $\ket{\psi^{(1)}(\tau_1)}=e^{-\frac{\mathrm{i}}{\hbar}\hat{H}^{(1)}\tau_1}\ket{\psi_0}^{(1)}$, $\ket{\psi_0}^{(1)}$ being the initial state of the system of $N$ particles, and $\hat{H}^{(1)}$ the Hamiltonian in this frame. The specific form of the initial state in terms of the kinematical space is not physically relevant - interested readers can find the details in Ref.~\cite{giacomini_SQRF}. The Hamiltonian is
\begin{align}\label{eq:QRF_hamiltonian_1}
    \hat H^{(1)} &= \hat \gamma_{\sum \mathbf{k},1}\left(\sum_i \sqrt{g'_{00}(\mathbf{q}_i,\mathbf{q}_L)}(c\hat \omega_{\mathbf k}^i + \hat \gamma_i^{-1} \hat H_i) + c \hat \omega_{\sum \mathbf{k},1} + \sqrt{g^{00}(\mathbf{q}_L)}\left(m_Lc^2 + \frac{\hat {\mathbf k}_L^2}{2m_L} + \hat H_L\right)\right) \nonumber \\
    &\sim \sum_{J=1}^L m_J c^2 + \sum_{j=2}^L \frac{\hat{\mathbf k}_j^2}{2m_j} +\sum_{J=1}^L\frac{(\sum_{j=2}^L \hat{\mathbf k}_j)^2}{2m_1^2}m_J + \sum_{i=2}^N m^{(i)} (V(\hat{\mathbf q}_i-\hat{\mathbf q}_L) - V(\hat{\mathbf q}_L)) - m_L V(\hat{\mathbf q}_L) + \nonumber\\ 
    &+ \sum_{i=2}^N\left(1+\frac{V(\hat{\mathbf q}_i- \hat{\mathbf q}_L) - V(\hat{\mathbf q}_L)}{c^2} + \frac{(\sum_{j=2}^L \hat{\mathbf k}_j)^2}{2m_1^2c^2} - \frac{\hat{\mathbf k}_i^2}{2m_i^2c^2}\right)\hat H_i + \left(1-\frac{V(\hat{\mathbf q}_L)}{c^2} + \frac{(\sum_{j=2}^L \hat{\mathbf k}_j)^2}{2m_1^2c^2} - \frac{\hat{\mathbf k}_L^2}{2m_L^2c^2}\right) \hat H_L. 
 \end{align}
Analogously, we can find the state associated to the laboratory, with the transformation operator
\begin{equation}\label{eq:QRF_transformation_operator_L}
     \mathcal{\hat T}_L = e^{\frac{i}{\hbar} \hat {\mathbf x}_L (\hat f^1 - \hat {\mathbf p}_L)}e^{\frac{i}{\hbar} \hat x_L^0(\hat f^0 - \hat p_L^0)}.
\end{equation}
The result is
\begin{align}
        \ket{\psi}^{(L)} &= \bra{q_L=0} \mathcal{\hat T}_L \ket{\Psi}_{ph} =\nonumber \\
        &= \int d\tau_L e^{-\frac{i}{\hbar}\hat H^{(L)}\tau_L}\ket{\psi_0^{(L)}}\ket{\tau_L}\label{eq:definition_state_L},
\end{align}
where the Hamiltonian from the perspective of $L$ is
\begin{align}
    \hat H^{(L)} &= m_L c^2 + \frac{\left(\sum_I \hat{\mathbf  k}_I\right)^2}{2m_L} + \sum_{I=1}^N \Delta(\hat{\mathbf  q}_I,\hat{\mathbf k}_I)(\hat \omega_{\mathbf k}^I + \hat H_I) = \nonumber \\
    &\sim \sum_{J=1}^L m_J c^2 + \sum_{I=1}^N \frac{\hat{\mathbf k}_I^2}{2m_I} + \sum_{I=1}^N m_I V(\hat{\mathbf q}_I)  +\left(1+\frac{V(\hat{\mathbf q}_I)}{c^2} - \frac{\hat{\mathbf k}_I^2}{2m_I^2c^2}\right)\hat H_I.\label{eq:QRF_Hamiltonian_L}
\end{align}
The Hamiltonian of Eq.~\eqref{eq:QRF_Hamiltonian_L} is the standard Newtonian Hamiltonian for free particles with the gravitational time dilation of the clocks due to their relative distance to the Earth and the special relativistic time dilation, coinciding with the Hamiltonian used in Refs.~\cite{lammerzahl,zych,zychThesis,pikovski}. Therefore, the introduction of the constraint of Eq.~\eqref{eq:constraint_CL} provides a consistent way to describe the reference frame of the laboratory.

The formalism also allows to change between perspectives, say from the perspective of the laboratory to the perspective of a particle. This can be achieved via the invertible transformation
\begin{equation}\label{eq:QRF_QRF_transformation}
    \ket{\psi}^{(1)} = \bra{q_1=0} \mathcal{\hat T}_1 \mathcal{\hat T^\dagger}_L \ket{\psi}^{(L)}\ket{p_L=0}.
\end{equation}
\begin{figure}[t]
     \centering
     \begin{subfigure}{0.45\textwidth}
         \centering
         \includegraphics[width=\columnwidth]{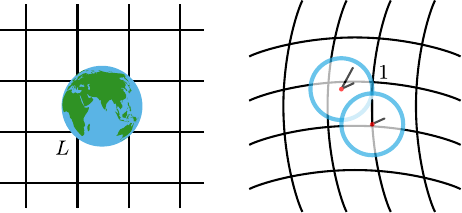}
         \caption{Perspective of $L$.}
         \label{fig:QRF_perspective_L}
     \end{subfigure}
     \hfill
     \begin{subfigure}{0.45\textwidth}
         \centering
         \includegraphics[width=\columnwidth]{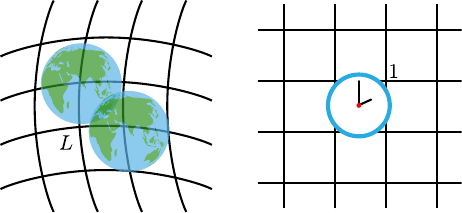}
         \caption{Perspective of 1.}
         \label{fig:QRF_perspective_1}
     \end{subfigure}
     \caption{Representation of an example of change of SQRF. In Figure~\ref{fig:QRF_perspective_L}, particle 1 is seen in a superposition of states and in a curved spacetime from the point of view of particle $L$, namely the laboratory. In Figure~\ref{fig:QRF_perspective_1} it is $L$ to be in a superposition and in a curved spacetime, when it is seen in the SQRF of particle 1.}
     \label{fig:QRF_change_perspective}
\end{figure}

A pictorial representation of this change of perspective is given in Figure~\ref{fig:QRF_change_perspective}. Note that the SQRF transformation to the perspective of a particle serve as an operational definition of its QLIF, since in that frame the metric is locally flat at the origin of the coordinates.

\subsection{Quantum extension of LPI}\label{sec:appa_QUGR}

In this section we report in further detail the calculations regarding the Quantum extension of LPI of Section~\ref{sec:QRF_perspective}. We employ the SQRFs framework described in Appendix~\ref{sec:SQRF}, using the extended constraints to include the laboratory as well. We focus on the Newtonian limit, namely we discard the special relativistic effects, including special relativistic time dilation, since we are only interested in gravitational time dilation.

We illustrate the simple case of a state in the perspective of the laboratory that is localised in position basis, namely
\begin{equation}\label{eq:QEP_initial_state_easy_case}
    \ket{\psi_0^{(L)}}_{\A \B} = \ket{x_1}_\Ae \ket{x_2}_\Be \ket{\tau_{in} =0}_\Ai \ket{\tau_{in}=0}_\Bi.
\end{equation}
This situation corresponds to having two quantum systems at two different heights $\mathbf{x}_{1,2}$, and to initialising the internal time of both of them is set to $\tau_{in}=0$. The particles are not in a superposition, so we do not expect to see any quantum effects. An arbitrary quantum state, however, can be obtained by linear combination of this simple state.

The state in the perspective of particle $A$ can be obtained with the reference frame transformation defined in Eq.~\eqref{eq:QRF_QRF_transformation}, namely
\begin{equation}
    \ket{\psi}^{(A)} = {}_\Ae \bra{q_A = 0} \mathcal{\hat T}_A \mathcal{\hat T}_L^\dagger \int d\tau_L e^{-\frac{i}{\hbar}\hat H^{(L)}\tau_L} \ket{\psi_0^{(L)}}_{\A \B} \ket{\tau_L}_\Li \ket{p_L=0}_L,
\end{equation}
where $\mathcal{\hat T}_A$ and $\mathcal{\hat T}_L$ are defined in Eq.~\eqref{eq:QRF_transformation_operator_1} and Eq.~\eqref{eq:QRF_transformation_operator_L} respectively, and the Hamiltonian $\hat H^{(L)}$ is defined in Eq.~\eqref{eq:QRF_Hamiltonian_L}.

The calculations can be performed by letting the operator $\mathcal{\hat T}_A \mathcal{\hat T}_L^\dagger$ act on the Hamiltonian, and expanding both $\bra{q_A=0}$ and $\ket{p_L=0}$ with a Fourier series. Then, we calculate the action of $\mathcal{\hat T}_A \mathcal{\hat T}_L^\dagger$ on the states explicitly. The result is that the state in the perspective of particle $A$ at time $t_A$ is 
\begin{equation}\label{eq:QEP_state_atom_1_easy_case}
    \ket{\psi(t_A)}^{(A)}_{\B \L} = \ket{\Psi(t_A,x_1,x_2)}_{\Be \L} \Ket{\left(1+\frac{V(\mathbf{x}_2) - V(\mathbf{x}_1)}{c^2}\right) t_A}_\Bi,
\end{equation}
where $\Ket{\left(1+\frac{V(\mathbf{x}_2) - V(\mathbf{x}_1)}{c^2}\right) t_A}_\Bi$ is the clock of the second particle $B$, while its external state is contained in $\ket{\Psi(t_A,x_1,x_2)}_{\Be \L}$, along with the complete state of the laboratory:
\begin{align}
    \ket{\Psi(t_A,x_1,x_2)}_{\Be \L} &= \left(1-\frac{V(\mathbf{x}_1)}{c^2}\right)^2 e^{-\frac{i}{\hbar}\hat H_{ext}^\prime \left(\left(1-\frac{V(\mathbf{x}_1)}{c^2}\right) t_A + \frac{x_1^0}{c}\right)} \times \nonumber \\
    &\times \ket{\sqrt{g_{00}(\mathbf{x}_1)}(x_2^0-x_1^0),\mathbf{x}_2-\mathbf{x}_1}_\Be \ket{-\sqrt{g_{00}(\mathbf{x}_1)} x_1^0,-\mathbf{x}_1}_\Le \Ket{\left(1-\frac{V(\mathbf{x}_1)}{c^2}\right)t_A}_\Li, \label{eq:Psi_QUGR}
\end{align}
where
\begin{equation}
    \hat H^\prime_{ext} = \sum_{J=A, B, L} m_J c^2 + m_A V(\hat{\mathbf x}_L) + m_B V(\hat{\mathbf x}_B-\hat{\mathbf x}_L) + \sum_{j=B, L} \frac{\hat{\mathbf p}_j^2}{2m_j} + \frac{\left(\sum_{j=2,M}\hat{\mathbf p}_j\right)^2}{2m_A}.
\end{equation}

The result of Eq.~\eqref{eq:QEP_state_atom_1_easy_case} is the usual expression of the gravitational redshift: given two clocks localised at two different heights, the proper time measured by a clock is dilated according to the other clock by a factor $\nu^\prime = \left(1+\frac{\Delta V}{c^2}\right)\nu$, where $\nu$, $\nu^\prime$ are the frequencies of the first and second clock respectively.
\begin{figure}[t]
     \centering
     \begin{subfigure}{0.45\columnwidth}
         \centering
         \includegraphics[scale=1]{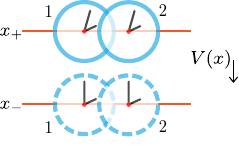}
         \caption{}
         \label{fig:QEP_initial_state_a}
     \end{subfigure}
     \hfill
     \begin{subfigure}{0.45\columnwidth}
         \centering
         \includegraphics[scale=1]{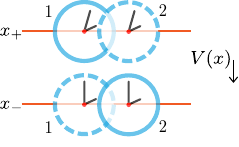}
         \caption{}
         \label{fig:QEP_initial_state_b}
     \end{subfigure}
     \caption{On the left, representation of two atomic clocks in a superposition in a gravitational field, where in each branch of the superposition they are at the same height. On the right, representation of two atomic clocks in a superposition in a gravitational field, where in each branch of the superposition they are at different heights. The first branch of the superposition is represented with solid lines, the second branch with dashed lines.}
\end{figure}
This result is employed in the main text to study Q-LPI when the clocks are entangled and follow the same path (Figure~\ref{fig:QEP_initial_state_a}). Here we give another example, namely the case in which in the laboratory frame the atoms are initially in the entangled state, represented in Figure~\ref{fig:QEP_initial_state_b},
\begin{equation}
    \ket{\psi_0^{(L)}}_{\A \B} = \frac{\ket{x_+}_\Ae \ket{x_-}_\Be + \ket{x_-}_\Ae \ket{x_+}_\Be}{\sqrt{2}}\ket{\tau_{in}}_\Ai\ket{\tau_{in}}_\Bi,
\end{equation}
where in each branch they are at different heights. The state in the perspective of particle $A$ can be calculated analogously to the previous case. We obtain a ``history state'' where now the clock on particle $B$ is time-dilated from the perspective of $A$, i.e., 
\begin{equation}
    \ket{\psi^{(A)}(\tau_A)}_{\B \L} = \frac{1}{\sqrt{2}}\left(\ket{\Psi(\tau_A,x_+,x_-)}_{\Be \L}\ket{\tau_A + \Delta \tau}_\Bi + \ket{\Psi(\tau_A,x_-,x_+)}_{\Be \L} \ket{\tau_A - \Delta \tau}_\Bi\right)
\end{equation}
where $\Delta \tau = \frac{V(\mathbf{x}_+) - V(\mathbf{x}_-)}{c^2}\tau_A$.

These results show that in the QRF of the clocks there is a superposition of gravitational time-dilations, which is due to the relative delocalisation of the clocks. Moreover, when the wavefunctions of the particles $A$ and $B$ have non-negligible uncertainties in position basis in each path, we find that, from the perspective of $A$, an additional effect of the spread of the wave function in each branch of the interferometer results in a redshift factor for each possible position of particle $B$ as seen from $A$. Therefore, the time of particle $B$ is dilated according to particle $A$, however the effect due to such a delocalisation is much smaller than that across different branches.

A similar procedure can be performed to generalise LLI to Q-LLI. Note that the transformation to the QLIF of $A$ does not map the momenta in the laboratory frame to the relative momenta with respect to $A$, due to the canonicity of the transformation. Hence, the functional relation between the relative velocities, defined as the time derivative of the position of $B$ in the QRF of $A$, and the momentum operator that is canonically conjugated to the position operator does not have a straightforward physical interpretation. However, the time of $B$ is not time-dilated as seen from $A$, and this holds true also for quantum superpositions of such states. Hence, we call this a generalization of LLI, namely Q-LLI.

\section{Calculations for the model for violations}\label{sec:appb}
In this appendix we give details on the calculations of the results presented in Section~\ref{sec:test_LZ}, showing what the violation of the EEP for QRFs entails in the perspective of one of the clocks.

The principle is formulated through the Eqs.~(\ref{eq:test_WEP_violation_condition}~-~\ref{eq:test_LPI_violation_condition})
\begin{align}
    m^{(i)} \hat{\mathbb{1}}+ \hat H^{(i)}(\hat{\mathbf p}) &\neq m^{(g)} \hat{\mathbb{1}} + \hat H^{(g)} (\hat{\mathbf x}) & (\text{Q-WEP}), \label{eq:Q_WEP_app}\\
    \hat H^{(r)} &\neq \hat H^{(i)} (\hat{\mathbf p}) & (\text{Q-LLI}), \label{eq:Q_LLI_app}\\
    \hat H^{(r)} &\neq \hat H^{(g)} (\hat{\mathbf x}) & (\text{Q-LPI}) \label{eq:Q_LPI_app},
\end{align}
where $i$ stands for ``inertial'', $g$ for ``gravitational'', and $r$ for ``rest''. It is easy to understand why this model can test a violation of the EEP for QRF by remembering that its formulation relies on a QRF transformation to a QLIF. In such a QLIF, the metric is locally minkowskian and, in turn, the behaviour of the clocks is independent of its position or velocity; hence, a violation of the EEP for QRFs implies a negation of this condition. We show in the following that this is achieved by imposing the previous equations, leading to a dependence of the time shown by the clock on the position or momentum operators. Clearly, different ways of violating the principle are possible, but because we are testing the violations employing the behaviour of quantum clocks, this is the most natural condition to impose.

Here, we investigate for the first time the effects of the violation not only in the perspective of the laboratory, but also in the QLIF of each particle. The result is that the violation of any condition is sufficient to break the SQRFs formalism, namely it prevents us to transform to the QLIF of a particle. Hence, the violations break the conditions for the validity of a description from a generic QLIF. Notice that it is necessary, in order to perform the test, that the perspective of the laboratory can be always described. In our model this is always true, as a consequence of imposing a Newtonian dynamical constraint for the laboratory, so we do not need to impose this condition separately. Intuitively, this happens because imposing a Newtonian constraint corresponds, as commented in Appendix~\ref{sec:SQRF}, to choosing a set of coordinates in which the metric is locally Minkowskian at the location of the laboratory. As a consequence, we cannot describe the frame of each particle as a QLIF, but we can do it for the laboratory: the laboratory identifies a preferred reference frame.

To keep the technical calculations simple, we investigate the violation of each condition (Eqs.~\ref{eq:test_WEP_violation_condition}~-~\ref{eq:test_LPI_violation_condition}) separately, choosing the appropriate set of constraints according to what aspect we are studying. However, this has no influence on the conceptual bearing of our results.

\subsection{Violation of Q-LPI}\label{sec:test_SQRF_LPI}

In this section we study the violation of LPI, formalised in Eq.~\eqref{eq:Q_LPI_app}. Since the condition involves only the gravitational Hamiltonian $\hat H^{(g)} (\hat{\mathbf x})$ and the rest Hamiltonian $\hat H^{(r)}$, without any dependence on the inertial Hamiltonian, Q-LPI does not depend on special relativistic effects. Thus, we study the violation of Q-LPI in the Newtonian limit of the SQRFs formalism, where the gravitational field is weak, and special relativistic effects are discarded. Hence we employ the slow-speed limit of the constraints of Eqs.~(\ref{eq:constraint_CI} -~\ref{eq:constraint_f1}).

To introduce the violation of Q-LPI, we modify the constraints distinguishing the gravitational Hamiltonian from the rest Hamiltonian for each particle, namely
\begin{equation}
    \hat H_I^{(g)}(\hat{\mathbf x}_I - \hat{\mathbf x}_L) \neq \hat H_I^{(r)}, \qquad I=A, B.
\end{equation}

The separation between $\hat H_I^{(g)}(\hat {\mathbf x}_I - \hat {\mathbf x}_L)$ and $\hat H_I^{(r)}$ is such that the gravitational Hamiltonian is coupled to the gravitational field, while the rest Hamiltonian is not. This results in a modification of the $\hat f^0$ constraint, namely
\begin{align}
    \hat f^0 &= \sum_{I=A, B}\left[ \hat p_I^0 + \frac{\hat H_I^{(r)}}{c} + \left(\sqrt{g_{00}(\hat {\mathbf x}_I - \hat {\mathbf x}_L)} -1 \right) \frac{\hat H_I^{(g)}(\hat {\mathbf x}_I - \hat {\mathbf x}_L)}{c}\right] + \frac{\hat H_L^{(r)}}{c}+ \hat p_L^0, \label{eq:test_constraint_f0_LPI}
\end{align}

indeed an expansion of the constraints for a weak gravitational field shows that $\hat H_I^{(g)}(\hat {\mathbf x}_I - \hat {\mathbf x}_L)$ is always coupled to the gravitational potential.

Note that it is not necessary to separate the Hamiltonians for the mass $L$, because the only Hamiltonian for $L$ that appears in the constraints is the rest Hamiltonian, since the metric at the location of $L$ is flat.

The state in the perspective of particle $A$ can be obtained via the usual prescription in Eq.~\eqref{eq:QRF_state_particle_1}, where the $\hat{\mathcal{T}}_A$ operator is defined as in Eq.~\eqref{eq:QRF_transformation_operator_1}, but $\hat f^0$ is replaced by the new constraint defined in Eq.~\eqref{eq:test_constraint_f0_LPI}. 

This results in a quantum state from the perspective of particle $A$ corresponding to
\begin{align}\label{eq:test_state_SQRF_LPI}
    \ket{\psi}^{(A)} = &\int d\tau_A d\mathbf{q}_L dt_A e^{-\frac{i}{\hbar} \hat{H}^{\prime (A)}\tau_A} \ket{\tilde \psi_0^{(A)}(\mathbf{q}_L,t_A)}_{\B \L}\times \nonumber \\ &\times e^{-\frac{i}{\hbar}\frac{V(\mathbf{q}_L)}{c^2}(\tau_A-t_A)\hat H_A^{(g)}(\mathbf{q}_L)}\Ket{\tau_A - (\tau_A-t_A)\frac{V(\mathbf{q}_L)}{c^2}}_\Ai,
\end{align}
where the relationship between $\ket{\tilde \psi_0^{(A)}(\mathbf{q}_L,t_A)}_{\B \L}$ and the standard initial state $\ket{\psi_0^{(A)}}_{\B \L}$ is
\begin{equation}
    \ket{\psi_0^{(A)}}_{\B \L} = \int dt_A d\mathbf{q}_L \ket{\tilde \psi_0^{(A)}(\mathbf{q}_L,t_A)}_{\B \L},
\end{equation}
and the Hamiltonian is
\begin{align}
    \hat H^{\prime (A)} &= \sum_{J=A, B, L} m_J c^2 + \sum_{j=B,L} \frac{\hat{\mathbf k}_j^2}{2m_j} +\frac{(\sum_{j=B, L} \hat{\mathbf k}_j)^2}{2m_A} + \nonumber \\
    &+ m_B (V(\hat{\mathbf q}_B-\hat{\mathbf q}_L) - V(\hat{\mathbf q}_L)) - m_L V(\hat{\mathbf q}_L) + \nonumber\\ 
    &+ \frac{V(\hat{\mathbf q}_B- \hat{\mathbf q}_L)}{c^2} \hat H_B^{(g)}(\hat{\mathbf q}_B- \hat{\mathbf q}_L) + \left(1-\frac{V(\hat{\mathbf q}_L)}{c^2}\right) \hat H_B^{(r)}+ \left(1-\frac{V(\hat{\mathbf q}_L)}{c^2}\right) \hat H_L^{(r)}. 
\end{align}
Differently to the standard state in the perspective of particle $A$ of Eq.~\eqref{eq:QRF_state_particle_1}, the state in Eq.~\eqref{eq:test_state_SQRF_LPI} is not a history state, since it cannot be written in the form of Eq.~\eqref{eq:QRF_state_particle_1}: this is the consequence of introducing a violation of Q-LPI. Therefore, it is not possible to identify the time measured by the first particle as its internal time, and consequently the SQRFs formalism does not hold anymore. In other words, the violation of Q-LPI implies that it is not possible to describe the system from the QLIF of particle $A$, and that Q-LPI is a necessary condition for the existence of a QLIF in the first place. 

Note that it is still possible to obtain the history state in the perspective of the laboratory, because the Hamiltonian of the mass $L$ is not modified by the violation of Q-LPI, since in the original constraints the metric for $L$ is flat, and thus in the modified constraint of Eq.~\eqref{eq:test_constraint_f0_LPI} the gravitational Hamiltonian $\hat H_L^{(g)}$ for the mass $L$ does not appear. 

The state in the perspective of the laboratory is
\begin{equation}
    \ket{\psi}^{(L)} = \int d\tau_L e^{-\frac{i}{\hbar} \hat H^{\prime (L)}\tau_L} \ket{\psi_0}_{\A \B}^{(L)} \ket{\tau_L}_\Li,
\end{equation}
which is still a history state. The modified Hamiltonian is
\begin{equation}\label{eq:test_H_lab_LPI}
    \hat H^{\prime (L)} = \sum_{J=A, B, L} m_J c^2 + \sum_{I=A, B} \left[ \frac{\hat{\mathbf k}_I^2}{2m_I} + m_I V(\hat{\mathbf q}_I) +\hat H_I^{(r)} + \frac{V(\hat{\mathbf q}_I)}{c^2} \hat H_I^{(g)}(\hat{\mathbf q}_I)\right].
\end{equation}
The fact that we are still able to give a description of the system in the perspective of the laboratory is not surprising, since the laboratory is a classical inertial frame in the Newtonian sense. This shows that the laboratory frame is a preferred frame, when Q-LPI is violated.

In the next sections we show that this holds true not only for violations of Q-LPI, but also for the entire EEP for QRFs. This allows us to predict the dynamics in the perspective of the laboratory, and in particular in the main text we calculate the probabilities of an interferometric experiment, using the most general Hamiltonian where also special relativistic effects are considered, namely without performing the slow-speed limit.

\subsection{Violation of Q-LLI}\label{sec:test_SQRF_LLI}

A test for Q-LLI does not involve gravitation, but only special relativity, since the Q-LLI violation condition in Eq.~\eqref{eq:Q_LLI_app} regards the inertial and rest mass, but not the gravitational mass. Therefore, we study the violation of Q-LLI in the special relativistic limit of the SQRFs formalism, obtained from the constraints of Eqs.~(\ref{eq:constraint_CI} -~\ref{eq:constraint_f1}) when gravity is neglected.

The constraints should be modified to account for the violation of Q-LLI, by distinguishing the inertial Hamiltonian from the rest Hamiltonian for each particle, namely
\begin{equation}
    \hat H_I^{(i)}(\hat{\mathbf p}_I) \neq \hat H_I^{(r)}, \qquad I=A, B.
\end{equation}
The separation is made in such a way that the inertial Hamiltonian is coupled to the momentum, while the rest Hamiltonian is not. The modified constraint is then
\begin{align}
    \hat f^0 &= \sum_{I=A, B}\left[ \hat p_I^0 + \left(\hat \gamma_I^{-1}-1 \right)\frac{\hat H_I^{(i)}(\hat{ \mathbf p}_I)}{c} + \frac{\hat H_I^{(r)}}{c}\right] +\hat p_L^0 + \frac{\hat H_L^{(r)}}{c},
\end{align}
where $\hat \gamma_I = \sqrt{1+\frac{\hat{\mathbf p}_I^2}{m_I^2c^2}}$ and $\hat \omega_{\mathbf p}^I = m_I c \ \hat \gamma_I$.

In order to obtain the state in the perspective of particle $A$, we employ the definition in Eq.~\eqref{eq:QRF_state_particle_1}, where the $\hat{\mathcal{T}}_A$ operator is
\begin{equation}
    \mathcal{\hat T}_A = e^{\frac{i}{\hbar}\hat {\mathbf x}_A ( \hat {\mathbf p}_B + \hat{\mathbf{p}}_L)} e^{\frac{i}{\hbar}\hat x_A^0 \left(\hat p_B^0 +\hat p_L^0 + \frac{\hat H_L^{(r)}}{c}+ \sum_{I=A, B} \left(\hat \gamma_I^{-1}-1 \right)\frac{\hat H_I^{(i)}(\hat{ \mathbf p}_I)}{c} + \frac{\hat H_I^{(r)}}{c}\right)}.
\end{equation}
Thus we find 
\begin{align}\label{eq:test_state_SQRF_LLI}
    \ket{\psi}^{(A)} = &\int d\tau_A d\mathbf{k}_B d\mathbf{k}_L dt_A e^{-\frac{i}{\hbar}\hat H^{\prime(A)}\tau_A} \ket{\tilde \psi_0(\mathbf{k}_B,\mathbf{k}_L, t_A)}^{(A)}_{\B \L} \times \nonumber \\ 
    &\times e^{-\frac{i}{\hbar}(1-\gamma^A_{\sum \mathbf{k}})(\tau_A-t_A)\hat H_A^{(i)} (\sum \mathbf{k})}\ket{t_A + \gamma^A_{\sum \mathbf{k}}(\tau_A-t_A)}_\Ai,
\end{align}
where $\hat \gamma^A_{\sum \mathbf{k}} = \sqrt{1+\frac{\left(\sum_{i=B,L} \hat{\mathbf k}_i\right)^2}{m_A^2c^2}}$, and we define for future convenience $\hat{ \omega_{\sum \mathbf{k}}^A} = m_A c \ \hat{\gamma_{\sum \mathbf{k}}^A}$. 

The relationship between the state $\ket{\tilde \psi_0(\mathbf{k}_B,\mathbf{k}_L, t_A)}^{(A)}_{\B \L}$ and the standard initial state $\ket{\psi_0}^{(A)}_{\B \L}$ is
\begin{equation}
    \ket{\psi_0}^{(A)}_{\B \L} = \int d\mathbf{k}_B d\mathbf{k}_L dt_A \ket{\psi_0(\mathbf{k}_B,\mathbf{k}_L, t_A)}^{(A)}_{\B \L},
\end{equation}
and the Hamiltonian is
\begin{equation}
    \hat H^{\prime (A)} = \hat \gamma_{\sum \mathbf{k}}^A \left(c \hat \omega_{\mathbf k}^B + m_L c^2 + \frac{\hat k_L^2}{2m_L} + c \hat \omega_{\sum \mathbf{k}}^A + \left(\hat \gamma_B^{-1}-1 \right)\frac{\hat H_B^{(i)}(\hat{ \mathbf k}_B)}{c} + \frac{\hat H_B^{(r)}}{c} + \frac{\hat H_L^{(r)}}{c}\right),
\end{equation}
The state in Eq.~\eqref{eq:test_state_SQRF_LLI} is not a history state, as a consequence of the violation of Q-LLI. Thus, if Q-LLI is violated, it is not possible to describe the system from the point of view of particle $A$, and the SQRFs formalism does not work anymore: a violation of Q-LLI prevents the existence of a QLIF for particle $A$. 

Notice that, in the presence of a violation, it is still possible to describe the laboratory, namely we still find a history state in the laboratory perspective, even if the QLIFs of the other particles do not exist anymore. The resulting Hamiltonian is
\begin{equation}\label{eq:test_H_lab_LLI}
    \hat H^{\prime (L)} = \sum_{I=A,B,L} m_I c^2 + \sum_{i=A,B} \left(c \hat \omega_{\mathbf k}^i + \left(\hat \gamma_i^{-1}-1 \right)\frac{\hat H_i^{(i)}(\hat{ \mathbf k}_i)}{c} + \frac{\hat H_i^{(r)}}{c}\right).
\end{equation}

\subsection{Violation of Q-WEP}\label{sec:test_SQRF_WEP}
The violation condition of Q-WEP in Eq.~\eqref{eq:Q_WEP_app} regards the inertial and gravitational mass, namely the parameters $m^{(g)}$ and $m^{(i)}$, as well as the operators $\hat H^{(g)}(\hat{\mathbf x})$, $\hat H^{(i)}(\hat {\mathbf p})$. Therefore, Q-WEP should be tested in the most general regime of the SQRFs formalism, where both general relativistic and special relativistic effects are significant. 

Nevertheless, the three aspects of the model for violations of the EEP are not independent. In fact, it is easy to show that the conditions in Eqs.~(\ref{eq:Q_WEP_app}~-~\ref{eq:Q_LPI_app}) imply e.g. that if Q-LPI is valid and $m^{(g)} = m^{(i)}$, then validity of Q-WEP coincides with validity of Q-LLI. Vice versa, if Q-LLI is valid and $m^{(g)} = m^{(i)}$, then Q-WEP coincides with Q-LLI. Therefore, we can reduce the test of Q-WEP to a test of Q-LLI and Q-LPI, that we already studied, and a test for the standard classical WEP, namely $m^{(g)} = m^{(i)}$.

Let us assume that Q-LLI is valid. Then, we can introduce a violation of Q-WEP in the SQRFs formalism with the conditions
\begin{align}
   & m_I^{(g)} \neq m_I^{(i)}, \hspace{2cm} I=A, B, \\
    &\hat H_I^{(g)}(\hat{\mathbf x}_I - \hat{\mathbf x}_L) \neq \hat H_I^{(r)}.
\end{align}
We already showed in Appendix~\ref{sec:test_SQRF_LPI} that the second condition prevents the SQRFs formalism to work. Therefore, a violation of WEP with this model implies that it is not possible to perform a transformation to the QLIF of a particle. Note that the standard classical WEP, namely $m_I^{(g)} = m_I^{(i)}$, $I=A, B$, does not play a role in the SQRFs mechanism.

This is not the most general violation of WEP possible, since we assumed Q-LLI to be valid. Nevertheless, even this weaker violation is sufficient to break the formalism, so the same argument holds for the general case of a stronger violation where Q-LLI is not assumed. Indeed, we can repeat the argument by assuming Q-LPI but not Q-LLI, obtaining the same result.

In conclusion, we have showed that if any of the three aspects of the EEP for QRFs is violated, the consequence is that it is not possible to associate a QLIF to a generic particle, and hence there are preferred frames.

\end{document}